\documentstyle[12pt,aasms4]{article}

\lefthead{K. Krisciunas}
\righthead{Mauna Kea Night-Sky Brightness}

\begin{document}

\title{Optical Night-Sky Brightness at Mauna Kea over the Course of a 
Complete Sunspot Cycle}
\author{K. Krisciunas\altaffilmark{1}}
\affil{Department of Astronomy, University of Washington, Box 351580,
Seattle, WA 98195}
\begin {center}
Electronic mail: kevin@astro.washington.edu
\end {center}

\altaffiltext{1}{Based on data obtained when the author was at the Joint 
Astronomy Centre, Hilo, Hawaii.}

\begin{abstract}
We have produced a data base of $V$-band and $B$-band night-sky brightness
measurements rather evenly spread out over the course of a whole sunspot
cycle from September 1985 to August 1996.  Almost all the data were obtained 
at the 2800-m level of Mauna Kea using the same telescope, same 
photomultiplier tube, filters, and diaphragm, thus minimizing various 
sources of systematic error and allowing an estimate of the sources of 
random error.  The yearly $V$-band averages of observed sky brightness 
ranged from 21.287 to 21.906 magnitudes per square arc second.  The color
of the sky is $B-V$ = 0.930 and does not change discernibly over the course
of the sunspot cycle.  After correcting the $V$-band data to the zenith,
we find that the airglow component varied a factor of 4.5 over the course
of the solar cycle. Once the 11-year solar cycle effect is removed from 
the data, the most significant contribution to the scatter of individual 
data points appears to be the short term variations on time scales of 
tens of minutes like those observed by the Whole Earth Telescope project.
\end{abstract}

\keywords{night-sky brightness}

\section{Introduction}
One of the most impressive sights in nature is the nighttime sky
unmarred by artificial light and with only a thin crescent Moon or 
no Moon in the sky.  On the Island of Hawaii
the air is often transparent enough that if one is fully dark adapted, the
starlight {\em alone} can cast a shadow of one's hand on one's chest,
if one is wearing a light colored coat.  This holds for clear, moonless
nights at sea level, at the top of Mauna Kea, or at elevations in between.
\footnote [2]{While this fact is impressive, it is not {\em that} strange if 
one considers that the starlight from the whole sky amounts to 1160 stars 
of first magnitude \markcite{ca73} (Allen 1973, p. 245).}

When doing astronomical photometry one wishes to determine the brightness
of an individual object such as a star.  One measures the star plus a
certain amount of surrounding sky, then must measure blank sky nearby
to obtain the light due to the star only.  Even in the absence of
sources of artificial light scattered into the beam,
the ``blank'' sky is not completely dark owing to four sources of light
\markcite{rg73} (Roach \& Gordon 1973): (1) zodiacal light and the
{\em gegenschein} (caused by sunlight scattered off interplanetary dust);
(2) faint unresolved stars and atomic processes within our Galaxy; (3) 
diffuse extragalactic light (due to distant, faint unresolved galaxies); 
and (4) airglow and aurorae (produced by photochemical reactions in the 
Earth's upper atmosphere).

\markcite {rg73} Roach \& Gordon (1973, p. 54) list some of the
reactions giving rise to airglow.  For example, continuum
radiation from 500 to 650 nm wavelength is emitted when NO$_{2}$ is 
produced from O and NO.  This light would be observable in the
Johnson $V$-- and $B$--bands.  An equally strong source of emission is the
line at 557.7 nm, due to [\ion {O}{1}].  This 
occurs in the middle of the Johnson $V$--band.  It has been known
for most of this century that the airglow varies over the course of the 
solar cycle.  (See \markcite {mw88} Walker 1988  for further information
and relevant references).  If we think of the airglow as a low-level aurora,
this makes sense.  As the solar wind energizes the Earth's atmosphere during
the day, detectable results can be observed at night.  

The airglow not only varies on times scales of 
years, but it systematically gets fainter over the course
of a given night (see Fig. 2 of \markcite{mw88} Walker 1988).  It can
even vary measurably over time scales of minutes (see Fig. 4 of 
\markcite{pil} Pilachowski et al. 1989, Fig. 3 of 
\markcite{ren} Nather et al. 1990, \markcite {mor} Morrison et al. 1997, and 
Fig. 5 below). As a result it is not sensible to speak 
of the ``intrinsic'' night-sky level at even a well protected site, except
in rather broad terms.  

While my main purpose over the years in doing photometry was to study
variable stars, it required very little extra effort to obtain the
data presented here.  The purpose of the present paper is to present further 
observations of the  night-sky brightness at Mauna Kea (primarily 
measured at the 2800-m level).  We now have data that are roughly evenly 
distributed over a complete solar cycle.  
These data are important for long term studies of the quality of the site
and for planning various observations.  As we all know, the signal-to-noise 
ratio obtained while imaging astronomical objects depends on the contrast of
the object itself and the sky background.  It is not well known, however,
that the optical sky brightness at a good site at solar maximum {\em with no 
Moon in the sky} is quite comparable to the sky brightness at solar 
minimum {\em with a quarter Moon in the sky}.  (See Krisciunas \& 
Schaefer 1991 for a model of the brightness of moonlight.)

In this paper we shall also consider the systematic and random errors that 
enter night-sky measurements.

\section{Observations}

The method of determining the night-sky brightness is laid out in 
Krisciunas (1990), hereafter referred to as Paper I, which contains
data of 1985 through 1989.
The principal system is described in Krisciunas (1996).  The 
telescope is a 15-cm f/5.82 reflector.  The photometer uses an uncooled
RCA 931A photomultiplier tube, a DC amplifier, and a strip chart recorder.
In short, a signal on the ``blank'' sky is made, with the strip chart
recorder gain set so as to obtain sufficient resolution for digitizing
the signal (typically 10 millivolts cm$^{-1}$ for the $V$-band and 5 
millivolts cm$^{-1}$
for the $B$-band).  One must also obtain dark current readings on the
same strip chart gains.  A nominal amplifier gain of 10$^{6}$ was always
used for the sky readings.  One or more standard stars were observed
near the zenith.  Standard stars were measured with the strip chart
recorder at a setting of 50 or 100 millivolts cm$^{-1}$, and with amplifier
gains on one of six settings between 10$^{3.5}$ to 10$^{6}$.  Since one
can determine a gain table for the amplifier in the lab and check for
errors with observations of standards stars of different brightness, we
feel that the gain settings are not a serious source of systematic error.
Still, for the purposes of accurately measuring sky brightness, it is 
best to observe both the sky and appropriately faint standard stars on the 
same amplifier gain setting.

Let D$_{sky}$ be the net deflection on the sky (i.e. with the dark current
subtracted off) and reduced to the same strip chart gain as the star 
measurement, D$_{\star}$ be the net deflection on a standard
star (i.e. with sky and dark current subtracted off), $\Delta$G be the 
difference of amplifier gain (in magnitudes) between the readings
on the sky and star, k$_{\lambda}$ be the atmospheric extinction for
the filter used (in magnitudes per air mass), X$_{\star}$ be the air mass 
value of the standard star observed, and M$_{\star}$ be the catalog 
magnitude 
of that standard star.  Then the single-beam magnitude of the sky reading, 
made with reference to that standard star is:
 
\begin{equation}
S = - 2.5\  \rm {log}\ (D_{sky} / D_{\star}) + \Delta G + k_{\lambda} 
X_{\star} + M_{\star}\  , \end{equation}

\parindent = 0mm
where the logarithm is to the base 10.  If we let A be the area of the
beam in square arc seconds, then the sky brightness in magnitudes per
square arc second will be

\begin{equation}
I(\mu)\  =\  S\  +\  2.5\  \rm {log}\  A\  .
\end{equation}

Throughout this paper we shall use the symbol $\mu$ to mean units of
magnitudes per square arc second.  $\mu_{V}$ and $\mu_{B}$ refer to
sky brightness in magnitudes per square arc second in the $V$-band and 
$B$-band, respectively.

\parindent = 9 mm

For our system the beam area is
6.522 $\pm$ 0.184 square arc minutes, and 2.5 log A is 10.927 $\pm$ 0.031.
If two or three standard stars are used, one can then average the 
derived values.  If one is carrying out all-sky photometry, in which
one determines the extinction and transformation to the UBV system from
observations of, say, a dozen standards, then one can derive the single
beam night-sky magnitude by setting the air mass of the sky reading to zero.
As stated in Paper I and described by \markcite{mw70} Walker (1970), if
the {\em only} component of the the sky brightness were faint background 
stars, one would treat it like an observation of any other star.  But
because airglow takes place in the atmosphere, not outside it, for
sky brightness one must do something a bit different.

In Table 1 we give night-sky values from the years 1990 through 1996.
The data of 9 February 1992 were obtained at the 4205-m Mauna Kea summit 
with the University of Hawaii 0.6-m telescope and an Optec SSP-5 photometer
using a 23 arcsecond diameter beam.  All the other data were taken with
the system described above at the 2800-m level of Mauna Kea.  This is
the mid-level facility at Hale Pohaku (Hawaiian for ``house of stone''),
the location of the Onizuka Visitors' Center.

In Table 1 UT$_{V}$ is the Universal Time of the $V$-band measurement.
The $B$-band measurement was taken within 5 minutes of this time.  
$\Delta$T$_{twi}$ is the time since the end of astronomical twilight
(Sun elevation $-18^{\rm o}$).  $\alpha$, $\delta$, and Z, respectively, 
are the Right Ascension, Declination, and zenith angle of the sky patch.  
The solar radio flux values (``Observed, Series C'') are provided by the 
Dominion Radio Astrophysical Observatory (DRAO).  They correspond to the 
day {\em prior} to date of the given sky measurement.

Including the data given in Paper I, we have 84 usable values, on 79 
nights, of the $V$-band night-sky brightness in our data base.  This 
amounts to 78 values on 75 nights at Hale Pohaku,
the rest having been obtained at sea level or at the Mauna Kea summit.  
In Paper I we found no discernible difference in the zenith night-sky 
brightness obtained at these three light-pollution-free locations.  
However, because of the increasing number of lights at Hale Pohaku over
the years, even though the street lights in the parking lots are shielded,
the night-sky brightness at Hale Pohaku must now be undoubtedly a bit
brighter than it is at the summit.  Our data base also contains 
70 usable $B$-band measures on 67 nights.  The data are termed ``usable'' 
if obtained on clear moonless nights, and at galactic latitude
$|b| >$ 10$^{\rm o}$.  This excludes data taken when the Moon was just below
the horizon.

\notetoeditor{Table 1 goes here}

Table 2 gives the observed yearly averages of sky brightness obtained
from 1985 through 1996.  We can obtain an estimate of the internal 
error of an individual measurement by subtracting off the yearly averages
from the data and computing the Gaussian standard deviation of the 
resultant distribution.  For the $V$-band measures we obtain $\sigma_{V} = 
\pm$ 0.173 $\mu_{V}$, and for the $B$-band measures we obtain $\sigma_{B} = 
\pm$ 0.181 $\mu_{B}$.  The mean color of the sky is $<B-V>$ = 0.930 $\pm$ 
0.018 based on 70 points.  

\notetoeditor {Table 2 goes here}

Our data exhibit no statistically significant variations of sky color over 
the course of the sunspot cycle.  Perhaps there {\em is} a variation, 
but given the internal error of an individual sky color ($\pm$ 0.147 
mag), it might take a century to demonstrate any color variation, or at 
least many more points per year over the course of one whole solar cycle. 
As a rule of thumb we can state that one need not necessarily {\em measure} 
the $B$-band sky brightness on a given occasion to have the value, if one has 
measured the stronger signal due to the $V$-band sky brightness.

In Fig. 1 we plot the the values from Table 2.  For comparison we 
plot in Fig. 2 the monthly averages of the 10.7 cm solar flux obtained
from DRAO.

\section{Discussion}

Because magnitudes are a logarithmic unit, for further analysis 
of the $V$-band measures we need to convert the values
of I($\mu_{V}$) to some linear brightness unit.  In times past one
used the unusual unit S$_{10}$($V$), the number of 10th magnitude
stars per square degree.  More recent studies (e.g. \markcite {rg89} 
Garstang 1989, \markcite{bs90} Schaefer 1990) have settled on the 
nanoLambert (nL).  \markcite {ca73} Allen (1973, p. 26) states that 
the surface brightness equivalent of one m$_{V}$ = 0 star per square 
degree is 2.63 $\times$ 10$^{-6}$ Lamberts.  Since a 10th
magnitude star is exactly 10$^{4}$ times fainter than a 0 magnitude star,
S$_{10}$(V) = 1 is the same as 0.263 nL.
Let $a \equiv (100)^{0.2} \approx 2.51189$, and let $Q \equiv 10.0\  
+\  2.5\  \rm{log}\ (3600^{2})\ \approx 27.78151$.  Q would be the brightness
of a single square arc second of sky if a $V$ = 10 star were spread out over
one square degree and that were the only source of light.  The observed sky 
brightness, B$_{obs}$, in nL is then related to the $V$-band sky brightness, 
I($\mu_{V}$), in magnitudes per square arc second, as follows:

\begin{equation}
B_{obs}(nL)\ =\ 0.263\ a^{[Q\ -\ I(\mu_{V})]}\ .
\end{equation}

\parindent = 0 mm

It is also possible \markcite {rg89} (Garstang 1989, Equation 28), to relate
the $V$-band sky brightness in magnitudes per square arc second to the 
brightness $b_v$ of the sky, expressed in photons cm$^{-2}$ sec$^{-1}$
steradian$^{-1}$, as follows:

\begin{displaymath}
I(\mu_{V}) \; = \; 41.438 \; - \; 1.0857 \; \rm{ln} \; (b_V) \; .
\end{displaymath}

For the $B$-band the first coefficient is changed to 41.965 (Garstang 1989, 
Equation 39).  Future work might sensibly use photon units, since nL are 
strictly applicable only to the eye, only approximately valid for the 
$V$-band, and not applicable to $B$-band measures.

\parindent = 9 mm

The sky brightness is faintest at the zenith.  We use Equation 19 of
\markcite {bs90} Schaefer (1990) to correct the observed sky brightness
to the zenith value:

\begin{equation}
B_{zen}\ =\ B_{obs}\ /\  (1\ +\ Z^{2}/2\ )\ ,
\end{equation}

\parindent = 0 mm
where Z is the zenith angle of the sky patch in radians.  A more elaborate
relation (i.e. Equation 7 below) must be used for large zenith angles.

\parindent = 9 mm

In Fig. 3 we give the yearly averages of sky brightness converted to
nL as a function of the 10.7 cm solar flux on the dates when sky brightness
was measured.  We cannot account for the anomalous yearly average
of 1985.  Perhaps we were not careful enough to exclude visible
stars from the sky patch.  Perhaps there was extra light scattered off of 
aerosols or dust from the ongoing series of eruptions of Kilauea Volcano
on the Big Island.  These eruptions began 3 January 1983.
The anomalous point of 1993, however, is probably due to ash from the
Philippine volcano Mt. Pinatubo.  This ash reached Hawaii on 1 July 1991,
just in time to degrade the sky for our solar eclipse of 11 July.
It gave us yellowish-grey sky during the day for nearly two years, colored
sunsets, and increased optical wavelength extinction values at least through
the spring of 1993.  However, what is suspect about our hypothesis for the 
1993 point in Fig. 3 is that the data of the second half of 1991 and for 1992
are not above the line.  The highest point in Fig. 3, at B$_{zen}$ = 96.5 
nL, is for the year 1990.

The least-squares line, fit to the dots in Fig. 3 and weighted by the 
errors of the points, is:

\begin{equation}
B_{zen}\ =\ (47.021 \pm\  2.784)\ +\ (0.2019\ \pm\ 0.0222)\ \times (10.7\  
\rm {cm\  solar\  flux})\ . \end{equation} 

\parindent = 0 mm
From 1947 to 1988 the observed 10.7 cm solar flux ranged from 63.0 to
383.4 $\times 10^{-22}$ W/m$^{2}$/Hz \markcite {mw88} (Walker 1988).  The 
data of 1989 through 1996 do not appear to have exceeded these extrema.  
Taking the extrema of the observed daily solar flux,
the implied  range of B$_{zen}$ is 59.74 to 124.43 nL, or $21.89 >$ 
I($\mu_{V}$) $> 21.09$.  The corresponding $B$-band range would be 
$22.82 >$ I($\mu_{B}$) $> 22.02$.  One would expect this range of sky 
brightness at {\em any} of the world's choice astronomical sites.

\parindent = 9 mm

Given our yearly means of $V$-band sky brightness, corrected to the
zenith, which ranged from 58.0 to 96.5 nL, Equation 5 implies that the
airglow contribution ranged from 11.0 to 49.5 nL, a factor of 4.5.
Equation 5 also implies that without the airglow we would detect 47.0 
$\pm$ 2.8 nL
due to zodiacal light, faint stars in the beam, and diffuse extragalactic
light.  We can estimate that 49.4 S$_{10}$(V) = 13.0 nL is due to
stars $V$ = 13 and fainter (\markcite {ca73}Allen 1973, p. 245) $-$ stars
we could not avoid because they are not visible in the eyepiece
of our simple photometer with a 15-cm telescope.  \markcite {jat95} Tyson 
(1995) indicates that
the diffuse extragalactic light amounts to 0.53 S$_{10}$(V), or
0.14 nL.  Thus, roughly, we would estimate that the {\em average} amount
of zodiacal light we measured was 33.9 ($\pm$ 3 or more) nL, the 
equivalent of 129 $\pm$ 11 S$_{10}$(V).
Since our average sky patch was taken at the zenith 3h 34m after the end
of twilight at latitude +20$^{\rm o}$, this corresponds to an average 
ecliptic longitude 157$^{\rm o}$ east of the Sun.  The ecliptic latitude
of the zenith at Mauna Kea is $-3^{\rm o} < \beta < 43^{\rm o}$, depending 
on the time of night and the day of the year.  \markcite {rg73}
Roach \& Gordon (1973, p. 46) indicate that the zodiacal light contribution
at $\lambda - \lambda_{\odot} \approx 157^{\rm o}$ ranges from 171 
S$_{10}$(V) on the ecliptic to 105 S$_{10}$(V) at $\beta$ = 43$^{\rm o}$.  
The value at the mid-range, $\beta \approx 22^{\rm o}$,
is about 133 S$_{10}$(V), in excellent agreement with our value just
given.  This gives us further confidence that we have no serious source
of {\em systematic} error in the calibration of our sky brightness values.
(We do note, however, that \markcite{lev} Levasseur-Regourd \& Dumont (1980)
found 108 S$_{10}$(V) for the zodiacal light at our mean ecliptic 
coordinates.)

\markcite {rg89} Garstang's (1989) Equations 28 and 39 let us relate
the color in magnitudes (or in magnitudes per square arc second, 
I($\mu$)) of some night sky component to the photon fluxes in 
photons cm$^{-2}$ sec$^{-1}$ steradian$^{-1}$, as follows:

\begin{displaymath}
I (\mu_{B}) \; - \; I(\mu_{V}) \; = \; 0.527 \; - \; 1.0857 \; \rm{ln} \; 
\left( \frac {b_B} {b_V} \right) \; . 
\end{displaymath}

\parindent = 0 mm

For the total light of the night sky we find $B - V$ = 0.930, so the
$B$-band to $V$-band flux ratio is 0.690.  For the zodiacal light only,
which should be the color of the Sun, $B - V$ = 0.65, the photon ratio would
be 0.893.  This allows us to determine the $B$-band contribution of the
zodiacal light.  Assuming that the photographic star counts (\markcite 
{ca73}Allen 1973, p. 245) are equal to the $B$-band star counts, we can 
then derive the $B$-band flux due to the airglow only.  The $V$-band
airglow, which is due to continuum plus 557.7 nm [\ion{O}{1}] emission,
ranges from 1.2 $\times$ 10$^7$ (solar minimum) to  5.5 $\times$ 10$^7$ 
photons cm$^{-2}$ sec$^{-1}$ steradian$^{-1}$ (solar 
maximum). The $B$-band airglow, which is essentially all continuum,
ranges from $<$ 0.6 $\times$ 10$^7$ photons 
cm$^{-2}$ sec$^{-1}$ steradian$^{-1}$ at solar minimum to 2.8 $\times$
10$^7$ at solar maximum. The $V$-band airglow intensity 
therefore increases by roughly
twice the $B$-band increase (4.3 $\times$ 10$^7$ vs. $>$ 2.2 $\times$ 10$^7$ 
photons cm$^{-2}$ sec$^{-1}$ steradian$^{-1}$) over the course of the 
solar cycle. The constancy of color provides a possibly useful restriction 
on the correlations between the airglow $V$ and $B$ intensities.

\parindent = 9 mm

\markcite {mw88} Walker (1988) found that the sky brightness on many 
nights decreased by 0.4 $\mu$ in the six hours
after the end of astronomical twilight.  To investigate whether our data
corroborates this, we subtracted the individual sky brightness values
(converted to nL and corrected to the zenith) from the regression line 
given by Equation 5.  The data set as a whole just gives a scattergram.
However, if we restrict ourselves to a subset of the data (1 September 1986
to 3 July 1987 and 20 December 1993 to 16 August 1996), which were 
obtained during solar minimum, we find marginal support for Walker's 
finding, which we show in Fig. 4.  Excluding the low point of 18 June 
1995 from the least-squares fit, we find

\begin{equation}
\Delta B_{zen}(nL)\ =\ (10.07\ \pm\ 3.24)\ -\ (1.79\ \pm\ 0.74)\ \times\ 
\Delta T_{twi}\ , \end{equation}

\parindent = 0 mm

where $\Delta$T$_{twi}$ is measured in decimal hours.

\parindent = 9 mm

The least-squares slope in Fig. 4, which excludes one
of the points, is only non-zero at the 2.4$-\sigma$ level.   The implied
nightly trend would be a diminution of the sky brightness by 10.76 nL
over the six hours since the end of twilight.  Given the average of
the individual data points (corrected to the zenith) that went into Fig. 4, 
this implies a sky brightness change from 69.63 to 58.87 nL, or I = 
21.72 $\rightarrow$ 21.91 $\mu_{V}$, about half the change found by Walker.  

Let us now consider the sources of uncertainty in a single sky brightness
measurement.  First, the sources of {\bf systematic} error:

\parindent = 0 mm

(1) Beam area.  This affects every derived value of sky brightness the
same way (see Equation 2).  Our 1$-\sigma$ uncertainty contributes 
$\pm$ 0.031 $\mu$ to the sky brightness values.

(2) Gain difference of amplifier ($\Delta$G in Equation 1).  This 
uncertainty is zero if the sky is observed on the same amplifier gain 
setting as the standard star, and is probably less than 0.01 mag for
the most common gain settings used on stars ($10^{5.0}$ and $10^{5.5}$).

(3) Gain difference of strip chart settings.  We assume this to be
negligible for the Hewlett Packard unit that we used, but it is not
necessarily zero.

(4) The average amount of faint background stars in the beam.  
We assume that stars fainter than V = 19 would be very difficult for 
{\em anyone} with a reasonably large beam to avoid.  We 
can estimate from \markcite {ca73} Allen (1973, p. 245) that $V$ = 13 to
19 stars contribute 44.9 S$_{10}$(V) units, or 11.81 nL (on average), to 
the sky brightness.  If we could have avoided these stars altogether
by using a much more sophisticated telescope, a smaller beam and
locations carefully chosen from the Palomar Observatory Sky Survey,
our brightest observed yearly average (not corrected to the zenith) of 
21.287 $\mu_{V}$, would have been 21.42 $\mu_{V}$, and the faintest 
observed yearly average of 21.906 $\mu_{V}$ would have been
22.15 $\mu_{V}$.  Choosing sky patches devoid of
stars was the method used by \markcite {mva} Mattila et al. (1996) at La 
Silla, but their values still exhibit quite a lot of scatter, indicating that
the sources of random error (6) and/or (7) below are very significant.

\parindent = 9 mm

Now let us consider the sources of {\bf random} error:

\parindent = 0 mm

(1) Calculating the air mass value of the standard star.  For objects 
close to the horizon, calculating the path length of atmosphere along the 
line of sight is non-trivial, but 
within 60 degrees of the zenith the air mass is well approximated by the 
secant of the zenith angle. For our purposes here, except for calculational
errors (using the wrong coordinates of the star or site), this is not a
source of uncertainty.  \footnote[3]{We mention calculational errors because
of examples such as the ``first pulsar planet'' \markcite {lyne} (Lyne \& 
Bailes 1992), which resulted from insufficiently accurate coordinates of 
the object.}

(2) Uncertainty of the catalog magnitude of the standard star.  Typically
$\pm$ 0.01 mag.

(3) Digitizing the signal of the standard star.  On average $\pm$ 0.02 mag.

(4) Uncertainty in the adopted extinction.  From extinction values given
in Paper I, the internal error of the $V$-band extinction at Hale Pohaku
is estimated to be $\pm$ 0.03 magnitudes per air mass on nights when the 
extinction is actually measured, and $\pm$ 0.06 if we just estimate it 
from the long term
seasonal averages and a subjective estimate of the quality of the night.  
Since the observations (of 1987 to 1996) were made at a median zenith angle
of 15.4 degrees (for which air mass X = 1.037) with a standard star close
by on the sky, the uncertainty in the next to last term of Equation 1 is 
essentially the uncertainty in the extinction.

(5) Digitizing the sky brightness reading.  Certainly $\pm$ 0.05 mag,
probably more.

(6) The ``time-since-the-end-of-twilight'' effect found by \markcite
{mw88} Walker (1988).  The average uncertainty due to this effect at 
Mauna Kea is unknown, since we only find it marginally at some times near 
solar minimum.  

(7) Variations of sky brightness over the course of minutes (see 
Fig. 4 of \markcite{pil} Pilachowski et al. 1989, Fig. 3
of \markcite {ren} Nather et al. 1990, \markcite {mor} Morrison et al. 
1997, and the discussion below).  Nather 
(private communication) indicates that this {\em can} amount to a 100 
percent change over the course of an hour.

(8) Differing amounts of faint background stars in the beam.   If the
faint star background varies by 50 percent of its average value, or 
$\pm$ 6 nL, it would translate to $\pm$ 0.063 $\mu_{V}$
at 21.3 $\mu_{V}$, and to $\pm$ 0.110 $\mu_{V}$ at 21.9 $\mu_{V}$.

\parindent = 9 mm

The quadratic sum of the first four real sources of random error (0.01, 0.02,
0.06, and 0.05) would give an internal error of $\pm$ 0.081 mag, which is
to be compared with our observed internal error of a given $V$-band
sky brightness value of $\pm$ 0.173 $\mu_{V}$.  
Even if (5) above is $\pm$ 0.10 mag, the sources of random error (6), 
(7), and (8) above must contribute over half of the variance.  Their
contribution to the internal error of a single reading must amount to
at least $\pm$ 0.13 mag.

\markcite {mor} Morrison et al. (1997) and Morrison (private communication)
discuss short term variations in the $R$-band.  They find 50 percent 
changes in their photometric zero point on time scales of minutes and even
find sky brightness changes over very small angular scales (i.e. different
parts of the {\em same} CCD frame).  It is becoming clear that area 
photometry at the highest levels of precision must take into consideration
what the airglow variations really are $-$ low-level auroral activity.

To investigate further these short term changes of sky brightness, we 
obtained some data from the Whole Earth Telescope project.  This 
amounts to three-channel photometry (program star, check star, sky) 
in {\em white light} (i.e. with no filter). The data are from two 
observing runs at the Canada-France-Hawaii Telescope at Mauna Kea that ran 
from 31 March to 3 April and 20 to 24 May 1991 UT.  (This was during 
the last solar maximum.)  During both of these observing runs the Moon 
was waxing, so it was above the horizon in the evening, and set before dawn.  

The WET data amount to thousands of 10 second integrations.  The 
dark count was less than 10 per second.  (Given that the count rate on 
the sky amounted to thousands of counts per second, we can ignore the dark 
count.)  Given that the data are white light
counts, not in any photometric band, it would be difficult
to calibrate the WET sky brightness counts to values in nanoLamberts using 
observations of the stars.  We can use the net measurements on the check 
stars to determine the
atmospheric extinction for the white light observations.  (Given the 
quantum efficiency as a function of wavelength of the blue 
sensitive photomultiplier tubes, we expected, and found, extinctions 
comparable to those one would measure at Mauna Kea in the $U$- or $B$-bands.)

From the check star measures we could determine when the sky was photometric,
and could determine a value of the extinction each night.  For our purposes
here we needed to eliminate the data taken when the Moon was adding to
the sky counts.  For crescent and quarter Moons, one may include data 
obtained when the Moon's zenith angle is greater than 94$^{\rm o}$.  For a 
gibbous Moon the critical zenith angle is about 96$^{\rm o}$.  (It would 
likely be a still greater value for full Moon.)  

Because some of the data were taken at high air mass (3.1), to derive
the zenith counts from the observed counts one must do something 
different than is done above with Equation 4.  Instead, it is more 
appropriate to use

\begin{equation}
B_{zen}\ =\ B_{obs}\ 10^{+0.4\ k\ (X-1)}\ /\ X\ ,
\end{equation}

\parindent = 0 mm

where

\begin{equation}
X\ =\ (1\ -\ 0.96\ \rm{sin}^{2}Z)^{-0.5} 
\end{equation}

is the optical path length along a line of sight in units of air masses
(not quite the secant of the zenith angle), $k$ is the atmospheric 
extinction in magnitudes per air mass, and Z is the zenith angle.
Equations 7 and 8 are given by \markcite{ks91} Krisciunas \& Schaefer (1991)
and are based on the ideas of \markcite{rg89} Garstang (1989).  

\parindent = 9 mm

In Fig. 5 we show the sky counts (reduced to the zenith) from the WET data 
files on those portions
of six photometric nights when the moon was sufficiently below the horizon.
We cut off the data for each night 5 minutes before the onset of morning
astronomical twilight (defined to be when the Sun elevation angle 
is $-18^{\rm o}$).  Note that for 
plotting purposes the data for given nights have been offset.  Of particular
interest is the night of 31 March 1990.  The first part of the data on that
night was taken from 1.2 to 3.1 air masses, then a new field at 1.3
air masses was observed until the end of the night.  Given that the
two sets of data on that night knit together well, we have confidence that
the sky was photometric, we used an appropriate value of atmospheric 
extinction, and that Equations 7 and 8 above are appropriate for sky 
brightness measures that extend to high air masses.

Let us consider the relative change of the sky brightness by defining it as 
the slope of one of the linear portions of one of the subsections of 
data in Fig. 5 divided by the mean sky count for that subsection.  We find 
that the zenith sky counts often increase or decrease 5 to 8 percent per
hour, but the rate of change can be as high as 22 percent per 
hour. If we take the logarithms to the base 10 of the sky counts and then 
multiply them by 2.5, the corresponding sky brightness change is typically
between 0.05 and 0.09 mag h$^{-1}$, but it can be large as 0.24 mag h$^{-1}$
(i.e. on the night of 21 May 1991).

By contrast, we found above, from considerations such as those in Fig. 4,
a not very convincing sky brightness change of 2.8 percent per hour,
or 0.030 mag h$^{-1}$.  \markcite {mw88} Walker (1988) found an
exponential, not linear, nightly sky brightness effect which was primarily
limited to the first 3 hours after the end of evening astronomical
twilight.  His average sky brightness change amounted to 0.07 mag h$^{-1}$.
More recently \markcite {rg97} Garstang (1997) has reanalyzed Walker's
data, subtracting the estimated contributions due to light pollution, 
zodiacal light, and faint stars in the beam.  This eliminated the sharp
curvature during the first two hours of the night, but still showed a slow
decline ($\approx$ 3.7 percent per hour) with scatter similar to our Fig. 4.

On 3 of the 6 nights shown in Fig. 5 the sky brightness started to 
{\em increase} half an hour to an hour before the start of astronomical 
twilight.  Could this be due to the {\em reverse} of Walker's night sky 
brightness effect, namely, the energization of the upper atmosphere  
owing the Sun's light from the {\em coming} day? \footnote [4] {If we 
consider that the typical emission height of the airglow is 90 km above
the surface of the Earth (\markcite {rg73}Roach \& Gordon 1973, p. 54),
this seems unlikely.  It would require ``currents'' from the morning
terminator to be moving west at roughly twice the rotational
speed of the Earth.} Or is this just part of the random variations of the 
sky brightness?

Finally, we would wish to know if there is any periodicity in the
short term variations of sky brightness.  Consider that the Earth has a
size (but hardly the mass!) comparable to that of a white dwarf star, and if 
the variations of the sky brightness are due to gravity-mode oscillations 
in the Earth's atmosphere, they might occur on time scales comparable 
(i.e. tens of minutes) to the $g$-mode oscillations of white dwarfs.
One might also attempt to correlate sky brightness changes with the 
twice-daily tides. From power spectrum analysis of the WET sky counts, we 
find no evidence  for any periodicities in the data.  However, long 
stretches of WET data  might reveal power at some frequencies, {\em 
especially} at 1 cycle per day if Walker's nightly sky brightness effect 
regularly takes place.

\section{Conclusions}

We have shown that the zenith $V$-band sky brightness at the mid-level of 
Mauna Kea varies between 21.3 to 21.9 magnitudes per square arc second
over the course of a complete sunspot cycle.  Once the solar cycle 
effect is removed from the data, the largest contribution to the 
internal scatter of individual measurements of sky brightness would 
appear to be the short term variations on time scales of tens of 
minutes.  These short term variations have never been studied in detail.
Power spectra of some Whole Earth Telescope data presented here indicate 
that the variations are random, not periodic. 

From the data with our system and references given above we find that
the zenith $V$-band sky brightness breaks down as follows: (1) airglow, 
11.0 to 49.5 nL, varying quite smoothly over the course of the 11-year
solar cycle; (2) mean zodiacal light, 33.9 nL; (3) mean background 
stars, 13.0 nL; and (4) extragalactic background light, 0.14 nL.
(To convert these numbers to units of S$_{10}$(V), divide by 0.263.)
For comparison, a 4 day old Moon at 60$^{\rm o}$ zenith angle 
would contribute 29 nL to the zenith sky brightness, while a quarter Moon 
would contribute 99 nL (\markcite {ks91}Krisciunas \& Schaefer 1991, 
Table 2).  (This assumes a $V$-band extinction of 0.172 magnitudes per
air mass.) With telescopes that can be accurately set to pre-selected 
``blank'' fields, the background star contribution can be largely 
eliminated from one's measures.

In spite of these variations of sky brightness at the 2800-m level of 
Mauna Kea, Hale Pohaku is a very dark site.  In practical terms what 
would these observing conditions mean for the amateur astronomer (i.e. 
observing with the eye as the detector)?  At such a location as Hale 
Pohaku one can easily
detect the nearest quasar, 3C 273 (m$_{V}$ = 12.8) with a 15-cm telescope.  
While Pluto has been closer to the Sun than Neptune these past years, it
too was detectable in a 15-cm f/6 reflector.  On 19 June and 10 July 
1988 I easily found Pluto, at m$_{v}$ = 13.7,  well within my personal 
record of m$_{v}$ = 14.5 obtained with the AAVSO chart of the field of 
R Her on 2 June 1986.  Due to the solar cycle effect, one cannot reach one's 
limit in any given year, but suffice it to say that the Onizuka Visitors' 
Center at Hale Pohaku is one of the premier locations available to amateur
astronomers.

Since the optical sky brightness at the Mauna Kea summit is essentially
the same as that at Hale Pohaku, the sky brightness at Mauna Kea and at
other light-pollution-free sites will vary by a significant amount on an
11-year time scale.  Professional astronomers using expensive telescopes
must take into account the solar cycle effect when planning their observing 
runs on the faintest targets of opportunity and judging the efficacy of 
their state-of-the-art instruments.  Simple data, such as those 
presented in this paper, are important for providing a frame of reference
for the ongoing analysis of the quality of astronomical sites, and for
ensuring the preservation of the high quality of those sites.

\acknowledgments

I would like to thank Don Hall, the Director of the University of
Hawaii's Institute for Astronomy, and the late Tom Krieger, former
head of Mauna Kea Support Services, for making small telescope observing 
possible at Hale Pohaku (with electric power and parking lot lights one can 
turn off).  Tony Tyson provided useful insights concerning the extragalactic
background light. I thank Ed Nather and Gerald Handler for information 
relating to the Whole Earth Telescope project, and would particularly 
like to thank Travis Metcalfe for providing some WET data to work on.
Roy Garstang made a number of very constructive suggestions which improved
this paper.  Heather Morrison provided additional insights.

\newpage 

\figcaption {Yearly averages of night-sky brightness from Table 2 for
the $V$-band (top set of points) and the $B$-band (bottom points).}

\figcaption {Monthly averages of 10.7 cm solar flux from DRAO.}

\figcaption {Yearly averages of $V$-band night-sky brightness, 1985 
to 1996, converted to units of nanoLamberts, and corrected to the 
zenith.  The abscissa
values are the averages of the 10.7 cm solar flux on those days when
the sky brightness values were obtained.  The open triangle corresponds
to data of 1985, the open circle to data of 1993.  The linear fit is
to the dots, weighted by the errors of those points.}

\figcaption {The residuals of individual $V$-band data points from the
least-squares line of Equation 5, vs. the time since the end of
astronomical twilight.  Data of 1 September 1986 to 3 July 1987 and
data of 20 December 1993 to 16 August 1996 are shown.  The datum
of 18 June 1995 (open square) is not included in the linear fit.}

\figcaption {Uncalibrated 10 second white light integrations on the sky 
with a blue sensitive photomultiplier tube, obtained during two Whole
Earth Telescope runs at the Canada-France-Hawaii Telescope at Mauna Kea.
The data have been corrected to the zenith via Equations 7 and 8. From
bottom to top the data sets have been offset by 0, 0, 5000, 8000,
13000, and 13000 counts.  The night of 20 May 1991 did have some cirrus;
we include the sky counts at those times when the check star counts
were a reasonably smooth function of time.}

\begin{deluxetable}{rrrrrrrccc}
\tablewidth{0pc}
\tablecaption{Individual Observations of Night-Sky Brightness}
\tablehead{
\colhead{UT date} & \colhead{UT$_{V}$} &
\colhead{$\Delta$T$_{twi}$}        & \colhead{$\alpha$} &
\colhead{$\delta$}  & \colhead{Z} &
\colhead{V}         & \colhead{B} &
\colhead{10.7 cm $\odot$ flux}  \\
\colhead{}          & \colhead{hh:mm} &
\colhead{hh:mm}     & \colhead{hh:mm} &
\colhead{$^{\rm o}$}    & \colhead{$^{\rm o}$} &
\colhead{mag/sec$^{2}$} & \colhead{mag/sec$^{2}$} &
\colhead{10$^{-22}$ W/m$^{2}$/Hz} }

\startdata
1990   &       &       &       &      &      &        &        &       \nl
21 Mar &  9:23 &  3:37 & 11:14 & 19.0 &  4.4 & 21.489 & 22.322 & 225.7 \nl
28 Mar &  8:09 &  2:21 & 10:25 & 19.8 &  2.8 & 21.492 & 22.391 & 216.0 \nl
18 Apr &  6:48 &  0:52 &  9:28 & 20.0 & 10.0 & 21.235 & 22.204 & 234.9 \nl
17 Sep &  8:06 &  1:28 & 21:10 & 19.0 &  4.3 & 21.079 & 21.922 & 203.0 \nl
14 Nov & 10:58 &  5:59 &  5:07 & 19.0 & 13.7 & 21.590 & 22.340 & 185.3 \nl
 \nl
1991   &       &       &       &      &      &        &        &       \nl
 4 Jan &  6:22 &  1:09 &  4:29 & 19.2 & 22.4 & 21.444 & 22.315 & 175.8 \nl
 7 Mar &  8:10 &  2:28 & 10:00 & 20.0 & 17.3 & 21.428 & 22.423 & 209.9 \nl
 4 Apr &  7:25 &  1:34 &  8:44 & 18.2 & 16.1 & 21.209 & 22.400 & 195.9 \nl
12 May &  9:02 &  2:53 & 15:41 & 28.9 & 24.9 & 21.588 & 22.641 & 229.8 \nl
 9 Sep &  7:16 &  1:29 & 17:27 & 26.1 & 37.1 & 21.665 & 22.841 & 196.1 \nl
15 Sep & 10:23 &  4:43 &  0:39 & 20.5 & 14.7 & 21.247 & 22.300 & 181.3 \nl
30 Sep &  7:48 &  2:23 & 20:40 & 19.5 & 20.1 & 21.484 & 22.543 & 194.7 \nl
13 Oct & 10:51 &  5:37 &  1:05 & 20.0 & 22.0 & 20.968 & 21.892 & 188.0 \nl
10 Nov &  9:52 &  4:52 &  3:44 & 20.0 & 13.5 & 21.192 & 22.361 & 197.4 \nl
27 Dec &  8:11 &  3:01 &  5:29 & 12.9 & 30.2 & 21.536 & \nodata & 260.8 \nl
30 Dec &  7:43 &  2:22 &  4:30 & 19.2 & 30.7 & 21.441 & 22.329 & 254.4 \nl
\nl
1992   &       &       &       &      &      &        &        &       \nl
27 Jan &  7:38 &  2:11 &  4:28 & 19.2 & 19.8 & 21.418 & 22.218 & 209.0 \nl
31 Jan & 10:08 &  4:39 &  7:28 & 27.8 & 15.4 & 21.370 & 22.284 & 280.3 \nl
 3 Feb & 11:25 &  5:55 &  8:45 & 18.7 & 16.6 & 21.082 & 21.944 & 288.3 \nl
 9 Feb$^{a}$ & 12:48 &  7:16 & 10:24 & 33.7 & 22.2 & 21.252 & 22.234 & 
225.0 \nl
23 Mar &  7:45 &  2:08 & 10:17 & 19.8 & 11.7 & 21.312 & 22.138 & 160.7 \nl
26 Nov &  9:07 &  4:09 &  4:31 & 18.5 & 19.1 & 21.287 & 22.191 & 166.8 \nl
\nl
1993   &       &       &       &      &      &        &        &       \nl
25 Jan &  8:20 &  2:53 &  4:29 & 19.5 & 25.4 & 21.571 & 22.688 & 104.8 \nl
19 Feb &  7:17 &  1:40 &  4:30 & 19.4 & 33.4 & 21.537 & 22.361 & 126.0 \nl
24 Mar &  8:06 &  2:19 &  8:45 & 18.0 & 15.8 & 21.568 & 22.240 & 120.8 \nl
14 Apr &  6:57 &  1:03 &  8:45 & 18.4 & 19.0 & 21.207 & 22.144 &  97.2 \nl
23 Apr &  6:51 &  0:52 & 10:18 & 20.0 &  4.9 & 21.315 & 22.106 & 117.1 \nl
11 Nov &  8:48 &  3:49 &  2:49 & 29.2 & 16.7 & 21.416 & 22.326 &  90.1 \nl
20 Dec & 10:23 &  5:17 &  4:30 & 19.2 & 21.5 & 21.808 & 22.295 &  87.0 \nl
\nl
1994   &       &       &       &      &      &        &        &       \nl
 7 Mar &  7:12 &  1:30 &  5:41 &  9.3 & 32.7 & 21.612 & 22.280 &  95.5 \nl
 8 Apr &  7:07 &  1:14 & 10:24 & 33.7 & 15.9 & 21.846 & 22.778 &  72.8 \nl
\nl
1995   &       &       &       &      &      &        &        &       \nl
18 Jun &  8:44 &  2:16 & 17:39 & 12.6 & 23.3 & 22.123 & \nodata & 69.9 \nl
25 Jun &  9:28 &  3:00 & 17:36 & 12.0 & 28.9 & 21.739 & \nodata & 71.1 \nl
 3 Jul &  8:53 &  2:24 & 17:39 & 12.6 &  9.3 & 21.726 & 22.574 &  77.7 \nl
\nl
1996   &       &       &       &      &      &        &        &       \nl
15 Feb & 10:42 &  5:07 & 10:20 & 19.5 &  4.9 & 21.964 & 22.917 &  69.0 \nl
11 Mar &  8:31 &  2:47 & 10:20 & 19.5 & 12.6 & 21.742 & 22.735 &  68.9 \nl
21 Mar & 10:23 &  4:37 & 13:10 & 18.0 & 17.1 & 21.871 & \nodata & 69.3 \nl
23 Mar & 10:33 &  4:46 & 12:35 & 18.4 &  3.9 & 21.951 & 23.141 &  73.8 \nl
16 Aug & 14:15 &  8:08 &  0:57 & 23.4 &  9.3 & 22.001 & 22.686 &  67.6 
\enddata
\tablenotetext{a} {Data obtained at the 4205-m Mauna Kea summit with the 
0.6-m telescope.  All other data were obtained with a 15-cm telescope at the
2800-m level.}
\end{deluxetable}

\begin{deluxetable}{llrlr}
\tablewidth{0pc}
\tablecaption{Yearly Averages of Night-Sky Brightness}
\tablehead{
\colhead{Year}      & \colhead{V} &
\colhead{n$_{V}$}   & \colhead{B} &
\colhead{n$_{B}$}  \\
\colhead{}         & \colhead{mag/sec$^{2}$} &
\colhead{}         & \colhead{mag/sec$^{2}$} &
\colhead{}     }
\startdata
1985 & 21.507 $\pm$ 0.094 & 4 &  &  \nl
1986 & 21.769 $\pm$ 0.053 & 9 & 22.513 $\pm$ 0.025 & 4 \nl
1987 & 21.752 $\pm$ 0.090 & 7 & 22.803 $\pm$ 0.104 & 6 \nl
1988 & 21.616 $\pm$ 0.062 &11 & 22.498 $\pm$ 0.044 &11 \nl
1989 & 21.471 $\pm$ 0.043 &14 & 22.478 $\pm$ 0.049 &14 \nl
1990 & 21.377 $\pm$ 0.095 & 5 & 22.236 $\pm$ 0.084 & 5 \nl
1991 & 21.382 $\pm$ 0.062 &11 & 22.404 $\pm$ 0.079 &10 \nl
1992 & 21.287 $\pm$ 0.048 & 6 & 22.168 $\pm$ 0.049 & 6 \nl
1993 & 21.489 $\pm$ 0.074 & 7 & 22.309 $\pm$ 0.072 & 7 \nl
1994 & 21.729 $\pm$ 0.117 & 2 & 22.529 $\pm$ 0.249 & 2 \nl
1995 & 21.863 $\pm$ 0.130 & 3 & 22.574             & 1 \nl
1996 & 21.906 $\pm$ 0.046 & 5 & 22.870 $\pm$ 0.103 & 4 \nl             
\tablecomments{The night-sky measurements of 1987 to 1996 were made at a
median zenith angle of 15.4 degrees.  As a result, the values of
night-sky brightness at the zenith for those years would be, on average, 
0.039 $\mu$ fainter than the values in this table.}
\enddata
\end{deluxetable}

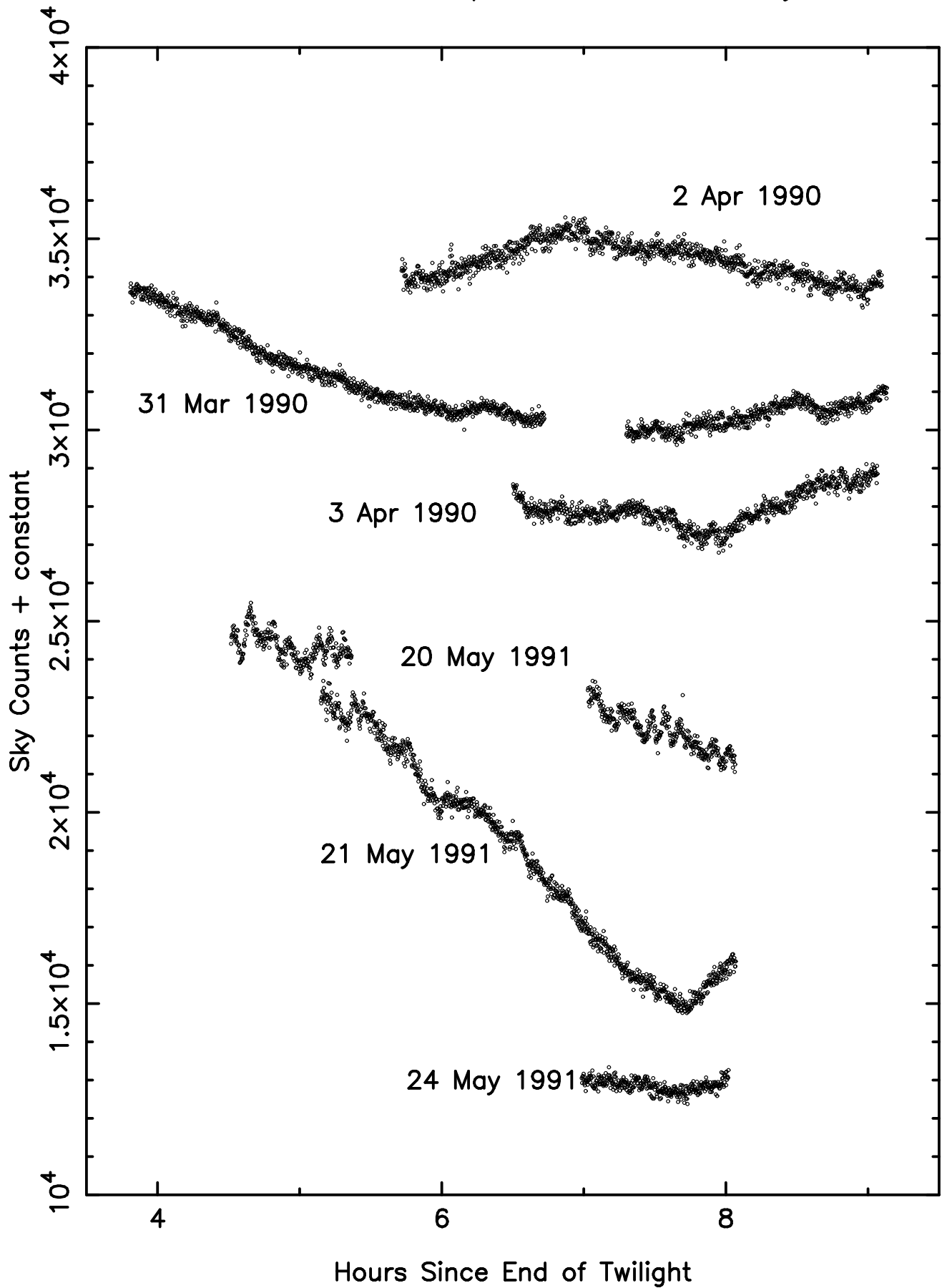
\end{document}